\documentclass[8.5pt,twoside,twocolumn]{article}
\usepackage[super,sort&compress,comma]{natbib}
\usepackage{mhchem}
\usepackage{times,mathptmx}
\usepackage{sectsty}
\usepackage{balance}

\usepackage{graphicx} 
\usepackage{lastpage}
\usepackage[format=plain,justification=raggedright,singlelinecheck=false,font=small,labelfont=bf,labelsep=space]{caption}
\usepackage{fancyhdr}
\usepackage{amsmath}

\def\v{{\bf v}}
\def\V{{\bf V}}
\def\w{{\bf w}}
\def\vs{{\bf s}}

\def\vr{{\bf r}}
\def\vR{{\bf R}}
\def\vrh{\hat{\bf r}}
\def\vsh{\hat{\bf s}}
\def\demi{ \frac{1}{2} }
\def\tLJ{t_{\textrm{LJ}}}
\def\tsim{t_{\textrm{sim}}}

\def\myw{0.5}

\begin{document}

\thispagestyle{plain}
\fancypagestyle{plain}{
\fancyhead[L]{\includegraphics[height=8pt]{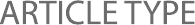}}
\fancyhead[C]{\hspace{-1cm}\includegraphics[height=20pt]{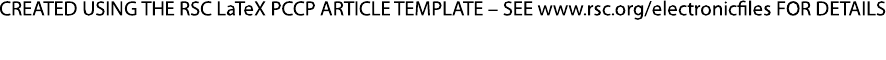}}
\fancyhead[R]{\includegraphics[height=10pt]{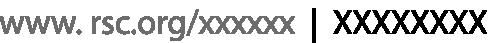}\vspace{-0.2cm}}
\renewcommand{\headrulewidth}{1pt}}
\renewcommand{\thefootnote}{\fnsymbol{footnote}}
\renewcommand\footnoterule{\vspace*{1pt}%
\hrule width 3.4in height 0.4pt \vspace*{5pt}}
\setcounter{secnumdepth}{5}

\makeatletter
\def\subsubsection{\@startsection{subsubsection}{3}{10pt}{-1.25ex plus -1ex minus -.1ex}{0ex plus 0ex}{\normalsize\bf}}
\def\paragraph{\@startsection{paragraph}{4}{10pt}{-1.25ex plus -1ex minus -.1ex}{0ex plus 0ex}{\normalsize\textit}}
\renewcommand\@biblabel[1]{#1}
\renewcommand\@makefntext[1]%
{\noindent\makebox[0pt][r]{\@thefnmark\,}#1}
\makeatother
\renewcommand{\figurename}{\small{Fig.}~}
\sectionfont{\large}
\subsectionfont{\normalsize}

\fancyfoot{}
\fancyfoot[LO,RE]{\vspace{-7pt}\includegraphics[height=9pt]{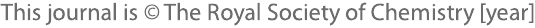}}
\fancyfoot[CO]{\vspace{-7.2pt}\hspace{12.2cm}\includegraphics{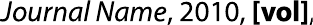}}
\fancyfoot[CE]{\vspace{-7.5pt}\hspace{-13.5cm}\includegraphics{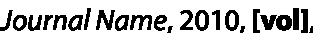}}
\fancyfoot[RO]{\footnotesize{\sffamily{1--\pageref{LastPage} ~\textbar  \hspace{2pt}\thepage}}}
\fancyfoot[LE]{\footnotesize{\sffamily{\thepage~\textbar\hspace{3.45cm} 1--\pageref{LastPage}}}}
\fancyhead{}
\renewcommand{\headrulewidth}{1pt}
\renewcommand{\footrulewidth}{1pt}
\setlength{\arrayrulewidth}{1pt}
\setlength{\columnsep}{6.5mm}
\setlength\bibsep{1pt}

\twocolumn[
  \begin{@twocolumnfalse}
\noindent\LARGE{\textbf{Phoretic self-propulsion: a mesoscopic description of reaction dynamics that powers motion$^\dag$}}
\vspace{0.6cm}

\noindent\large{\textbf{Pierre de Buyl,$^{\ast}$\textit{$^{a,b}$} and
Raymond Kapral\textit{$^{b}$}}}\vspace{0.5cm}

\noindent\textit{\small{\textbf{Received Xth XXXXXXXXXX 20XX, Accepted Xth XXXXXXXXX 20XX\newline
First published on the web Xth XXXXXXXXXX 200X}}}

\noindent \textbf{\small{DOI: 10.1039/b000000x}}
\vspace{0.6cm}

\noindent \normalsize{The fabrication of synthetic self-propelled particles and the experimental investigations of their
  dynamics have stimulated interest in self-generated phoretic effects that propel nano- and micron-scale objects. Theoretical
  modeling of these phenomena is often based on a continuum description of the solvent for different phoretic propulsion mechanisms, including, self-electrophoresis, self-diffusiophoresis and self-thermophoresis. The work in this paper considers various types of catalytic chemical reaction at the motor surface and in the bulk fluid that come into play in mesoscopic descriptions of the dynamics. The formulation is illustrated by developing the mesoscopic reaction dynamics for exothermic and dissociation reactions that are used to power motor motion. The results of simulations of the self-propelled dynamics of composite Janus particles by these mechanisms are presented.}  \vspace{0.5cm}
 \end{@twocolumnfalse}
  ]

\section{Introduction}
\footnotetext{\dag~Electronic Supplementary Information (ESI) available: [details of any supplementary information available should be included here]. See DOI: 10.1039/b000000x/}


\footnotetext{\textit{$^{a}$~Center for Nonlinear Phenomena and Complex Systems, Universit\'{e} libre de Bruxelles, Campus Plaine - CP231, 50 Av. F. Roosevelt, 1050 Brussels, Belgium. Fax: +32 2 650 57 67; Tel: +32 2 650 57 92; E-mail: pdebuyl@ulb.ac.be}}
\footnotetext{\textit{$^{b}$~Chemical Physics Theory Group, Department of Chemistry, University of Toronto, Toronto, ON M5S 3H6, Canada. %
E-mail: rkapral@chem.utoronto.ca; Fax: +1-416-9785325; %
Tel: +1-416-9786106}}


Nano- and micron-scale motors that are able to propel themselves
through solution by utilizing the energy derived from chemical
reactions constitute an interesting class of objects with numerous
potential applications. Such small self-propelled objects have been
made in the laboratory and their properties have been investigated. In
particular, a good deal of research has been devoted to some of the
earliest examples of motors of this type, bimetallic and striped
metallic rod
motors~\cite{paxton_et_al_nanorods_2004,ozin:05,sen:06,ozin-1:05,wang:06,wang:08}. Experimental
studies of Janus colloidal motors, where one face of a polystyrene or
silica sphere is coated with
platinum~\cite{howse_et_al_prl_2007,ke_et_al_jpc_a_2010}, and sphere
dimer motors, where two linked spheres, one silica and the other
platinum~\cite{valadares_el_al_sphere_dimers_small_2010}, have also
been carried out. Interest in such motors stems from their potential applications~\cite{mallouk:09,ozin-1:05,wang:09,pumera:10}.
For example, striped metallic rod motors that include Ni can be controlled by external magnetic fields and experiments have demonstrated that targeted cargo delivery can be achieved~\cite{sen:05-2,burdick:08,ghosh:09}, as has transport by Pt-Au nanomotors using photochemical stimuli~\cite{sen:10}. Self-propelled Janus particles have recently been investigated experimentally for cargo transport
applications~\cite{baraban_et_al_soft_matt_2012,baraban_et_al_acs_nano_2012}.

Such motors are placed in a solution containing fuel
of some type, most often hydrogen peroxide in the cases mentioned
above, and chemical reactions on portions of the motor give rise to
directed motion. In these examples the motors move by self-phoresis,
where the gradient of some field across the motor, which is generated
by asymmetrical chemical reactivity, induces fluid flow in the
surrounding medium resulting in propulsion. The metallic rod motors
exploit an electrochemical mechanism involving oxidation and reduction
reactions at the rod components, leading to self-electrophoresis~\cite{wang:06},
while the Janus and sphere dimer motors move by
self-diffusiophoresis. Motors that are propelled by the generation of
bubbles have also been constructed and studied~\cite{ismagilov:02,gibbs_zhao_appl_phys_lett_2009,manesh_et_al_acs_nano_2010}.
While many aspects of propulsion by self-phoresis are generic, the motor motion
depends on the nature of the chemical reactions taking place on the
motor, as well as those in the environment in which the motor moves.

A characteristic feature of such self-propelled motion is that the entire system
is force-free. This has implications for motor propulsion and for the nature of
the flows that develop in the system. The treatments of self-propulsion based on
the motion of colloidal particles in gradients rely on various continuum
approximations~\cite{anderson:84,anderson:89,anderson:82,anderson:91,julicher:09}. While
these approximations are certainly valid for large particles, and even for quite
small particles, as the nano and molecular levels are approached such
approximations should be subject to scrutiny. There have been numerous studies
of propulsion by self-phoretic
mechanisms~\cite{golestanian:05,golestanian:07,golestanian-1:07,popescu:09,popescu:10,sabass_seifert_jcp_2012}. In
common with our earlier studies of sphere dimer and polymer motors where a
simple reaction $A \to B$ occurs on the catalytic sphere~\cite{ruckner_kapral_prl_2007,yuguo:08,yuguo:09-2,snigdha:11a,snigdha:10,snigdha:11},
we consider a particle-based mesoscopic description of the dynamics of
self-propelled particles. The dynamics is such that the mass, momentum and
energy conservation laws are satisfied and hydrodynamic flows are correctly described.

In this article we show how the mesoscopic dynamics can be extended to treat a
general class of chemical reactions on the catalytic sites of the motor. The
reactive dynamics is constructed to satisfy the conservation laws so the
basic properties of the dynamical laws are preserved. As examples of these more
general reactions, we present the results of simulations of self-propulsion of a
composite Janus particle arising from a dissociation reaction $A \to 2B$ and an
exothermic reaction $A \to B +\Delta u$ at motor catalytic sites. These examples
illustrate some interesting effects; in particular, in the latter case of an
exothermic reaction propulsion occurs by self-thermophoresis. The general
strategy described in this paper should provide a framework that is useful in
explorations of the effects of complex chemical kinetics on self-propelled
motion at a mesoscopic level.

The outline of the paper is as follows: The system we consider comprises a motor
and the environment in which it moves. In Sec.~\ref{sec:meso} we describe the
motor itself, the (possibly reactive) environment in which it moves and how it
interacts with the chemical species in the environment. The detailed description
of the reactive processes at the motor catalytic sites is given in
Sec.~\ref{sec:chem-react} and the results of simulations of the motor dynamics
are presented in Sec.~\ref{sec:sim}. The conclusions of the paper are given in
Sec.~\ref{sec:conc}.

\section{Mesoscopic dynamical system}\label{sec:meso}

\begin{figure}[htbp]
  \centering
  \begin{minipage}[htbp]{.35\linewidth}
    \includegraphics[width=\linewidth]{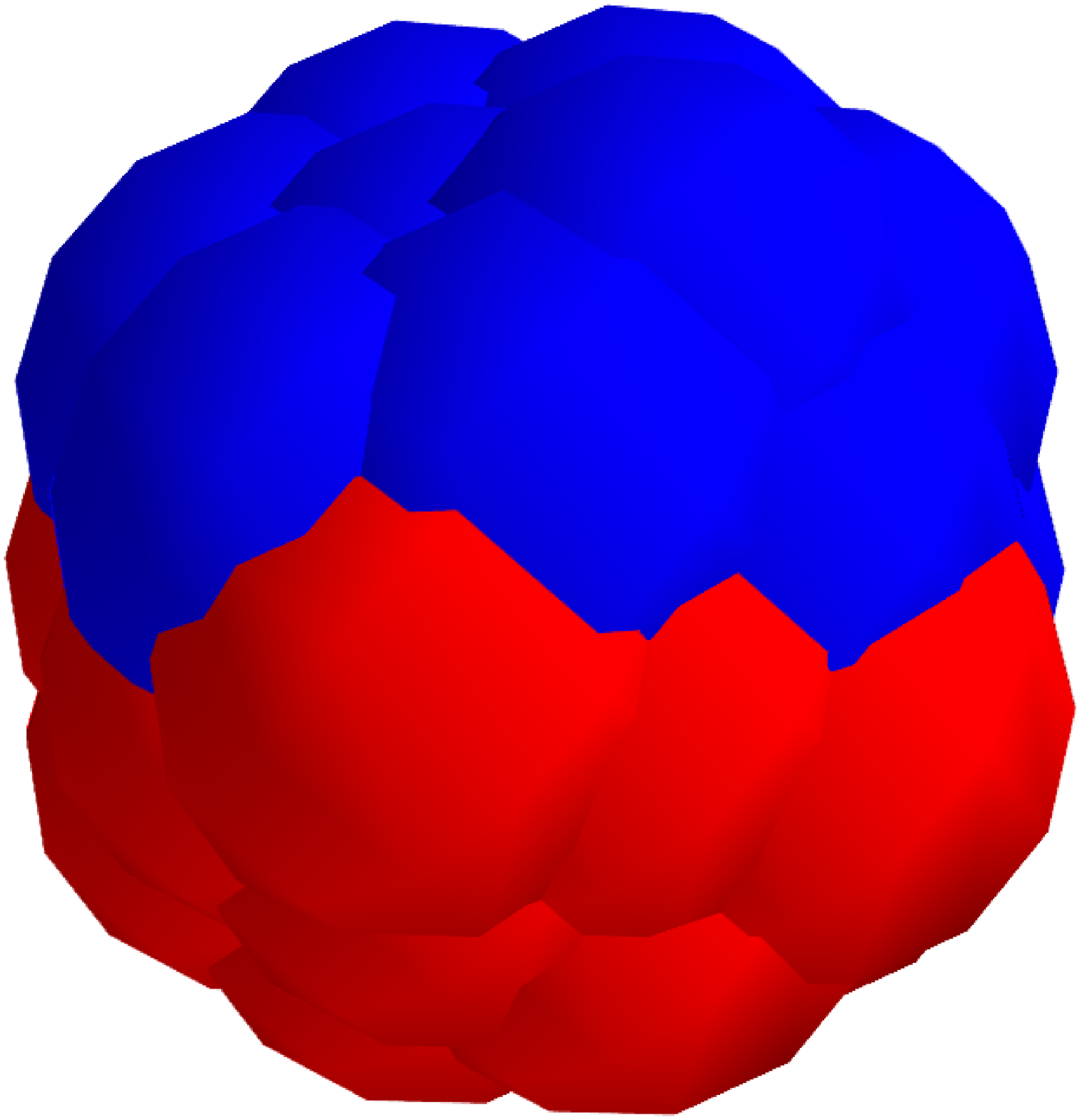}
  \end{minipage}
  \begin{minipage}[htbp]{.49\linewidth}
    \includegraphics[width=\linewidth]{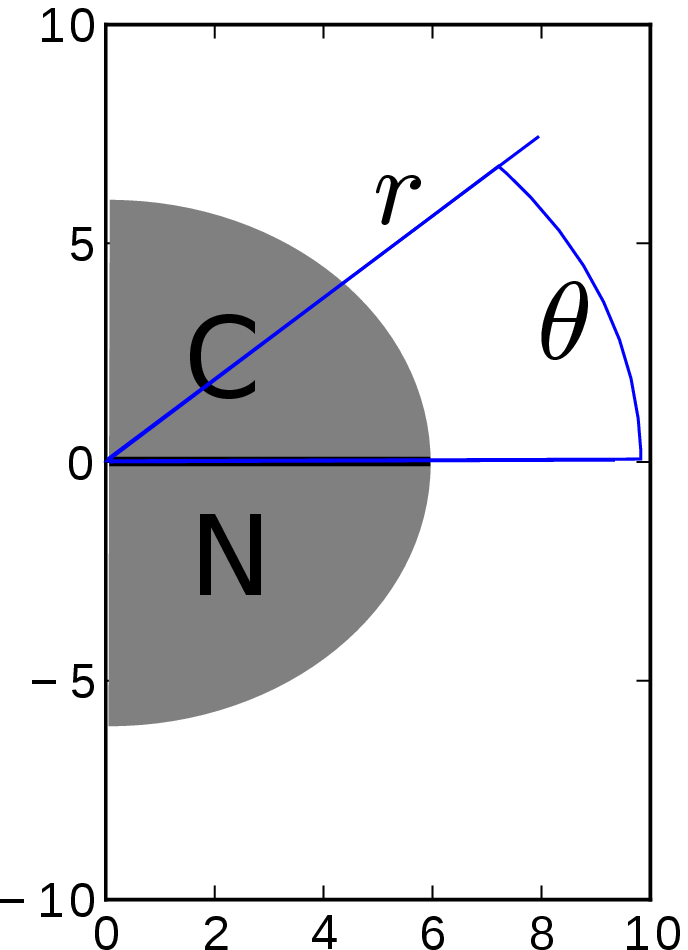}
  \end{minipage}
  \caption{Left panel: A representation of the bead model for the
    Janus particle. The top part is the catalytic side $C$ of the Janus particle
    and the bottom part is the noncatalytic side $N$. Right panel: The
    coordinates system defined with the Janus particle's center of mass as the
    center. The gray disk represents the Janus particle. In subsequent figures,
    that region contains empty data.}
  \label{fig:coord}
\end{figure}

The chemically-powered motor we study is a composite Janus particle comprising a
rigid assembly of catalytic and noncatalytic beads (see left panel of
Fig.~\ref{fig:coord}). One half of the Janus particle is made of catalytic ($C$)
beads while the other is half is made from non-active ($N$) beads. The solvent
in which the Janus particle moves contains, in general, reactive and nonreactive
chemical species. The beads in the Janus particle interact with solvent
particles through specific intermolecular forces. There are no explicit
solvent-solvent intermolecular forces. These are treated by multiparticle
collisions~\cite{malevanets_kapral_mpcd_1999,malevanets_kapral_mpcd_2000} as discussed below and described in reviews~\cite{kapral:08,gompper:09}.
In addition, the catalytic beads are able to catalyze various chemical reactions
that are responsible for the propulsion of the motor. It is the nature of these
chemical reactions that is the principal focus of this investigation and the
description of how general reactions are implemented on the catalytic sites will
be the topic of the next section. Here we specify the nature of the mesoscopic
dynamics that governs the evolution of the system.

The evolution of the entire system, the Janus particle and the surrounding
solvent is carried out by combining molecular dynamics (MD) with multiparticle
collision dynamics
(MPCD)~\cite{malevanets_kapral_mpcd_1999,malevanets_kapral_mpcd_2000}. In this
hybrid evolution scheme, Newton's equations of motion are integrated for all
interacting particles for time intervals $\tau$. This includes the Janus
particle and all solvent particles that interact with it. The remainder of
the noninteracting solvent particles free stream during these time intervals. At
the time intervals $\tau$ (possibly) reactive multiparticle collisions take
place as described in the next subsection.

\subsection{Dynamical model for the environment in which the motor moves}

The motor environment is modeled by reactive multiparticle collision dynamics
(RMPCD)~\cite{rohlf_et_al_rmpcd_2008}, since we wish to allow for the possibility
that chemical reactions may occur in the bulk of the
solution~\cite{snigdha:11a} in addition to the catalytic reactions that occur
on the motor itself. RMPCD combines multiparticle velocity-changing collisions
among solvent species with local probabilistic rules for changing the identities
of those solvent species that participate in reactive events in the bulk of the
solution. To carry out the dynamics, at the time intervals $\tau$ the system is
partitioned into cells labeled by $\xi$. We suppose that various reactions,
specified by the index $j$, occur in the solution:
\begin{equation}\label{eq:reactions}
\sum_{\alpha} \nu_{\alpha}^j  X_{\alpha} \mathrel{
\mathop{\kern0pt {\rightleftharpoons}}\limits^{{k_j}}_{k_{-j}}}
\sum_{\alpha} \bar{\nu}_{\alpha}^j  X_{\alpha}\;\;\;(j=1,\dots,r).
\end{equation}
The stoichiometric coefficients for reaction $j$ are $\nu^j_{\alpha}$ and
$\bar{\nu}^j_{\alpha}$ while $k_j$ and $k_{-j}$ are the rate constants for the
forward and reverse reactions. Each reaction, $\sum_{\alpha} \nu_{\alpha}^j
X_{\alpha} \mathrel{ \mathop{\kern0pt {\rightarrow}}\limits^{{k_j}}_{}}
\sum_{\alpha} \bar{\nu}_{\alpha}^j X_{\alpha}$, is taken to occur with
probability $p^\xi_j({\bf N}^\xi)=a_{j}^\xi \Big(1-e^{-a_0^\xi
  \tau}\Big)/a_0^\xi$, where ${\bf N}^\xi=(N_1^\xi,N_2^\xi,\dots,)$ is the set
of numbers of the different species in cell $\xi$, and $a_j^\xi = k_j(V_c)
\prod_{\alpha} N_{\alpha}^\xi!/(N_{\alpha}^\xi-\nu_{\alpha}^{j})! $ accounts for
the number of different ways the reaction can occur in the cell, with
$a_0^\xi=\sum_{j} a_j^\xi$. The rate constants $k_j(V_c)$ are scaled to account
for the cell volume $V_c$. Reaction rules may be constructed so that mass,
momentum and energy are conserved in the reactive events~\cite{rohlf_et_al_rmpcd_2008}.

Once the local reactive events have taken place, all particles in each cell
$\xi$ undergo multiparticle collisions where the post-collision velocity of
particle $i$ in cell $\xi$, $\v_i'$, is given
by~\cite{malevanets_kapral_mpcd_1999}
\begin{equation}
  \label{eq:mpcd}
  \v_i' = \V^\xi + \omega^\xi \cdot \left( \v_i - \V^\xi \right)
\end{equation}
where $\V^\xi$ is the center of mass velocity of all particles in cell $\xi$ and
$\omega^\xi$ is a rotation operator for the cell $\xi$. This algorithm preserves
mass, linear momentum and energy within each cell. The combined effect of
multiparticle collisions and reactive events is to mimic the effects of many
real reactive and nonreactive collisions that take place in the time
interval $\tau$. Since the full hybrid MD-RMPCD preserves the basic conservation
laws, hydrodynamic interactions and hydrodynamic flows induced by the
self-propulsion are properly described so that this mesoscopic simulation scheme
can capture all essential features of the Janus particle and solvent flow
motions.

\section{Chemical reactions at motor catalytic sites}\label{sec:chem-react}

In this section we show how different types of reaction can be implemented at a
mesoscopic level at the catalytic face of the Janus particle. As noted above, in
order to respect the basic conservation laws and insure that fluid flows are
properly described, the reactions must be constructed to satisfy mass, momentum
and energy conservation laws. In earlier studies on sphere dimer motors we
restricted our considerations to the simple thermoneutral catalytic reaction, $ A+C
\rightleftharpoons B+C$. This reaction amounts to a ``coloring'' process that
changes $A$ to $B$ on encounters with the catalytic $C$ sphere. Mass, momentum
and energy are simply conserved in this process. The propulsion arose from the
fact that the $A$ and $B$ species interacted with different intermolecular
potentials at the noncatalytic $N$ sphere.

The actual catalytic reactions that take place on motor active regions may be
quite complex, involving adsorption onto the surface with subsequent formation
of products and their release, or binding to active sites in enzymes followed by
conformational or other changes and product release. All of these events could
involve dissipation of energy into local solvent degrees of freedom, either
directly or subsequent to various internal energy transfer processes within the
motor. Our mesoscopic treatment of such reactive events coarse grains over most
of these molecular-level details. In particular, when the reagents lie within
the interfacial zone around the catalytic portion of the Janus particle where
intermolecular forces between the reagents and Janus particle beads are
non-zero, there is a possibility that a reaction will take place. We subsume all
of the reaction details into effective reactive events where reactive species
conversion and possible energy distribution in solvent degrees of freedom take
place just as the species exit the reactive interfacial zone. The local solvent
molecules that participate in any energy exchanges with reactive species are
taken to lie in the MPC collision cell that contains the reactive species.
Using this strategy it is possible to construct a general class of catalytic
reactive events on the active face of the Janus particle~\cite{footnote:enzyme}.

In this coarse-grain description of real reactions catalyzed on the Janus motor,
the conservation of momentum and energy in a general reactive event can be
written as
\begin{equation}
  \label{eq:mom}
  m_J \v_J + \sum_{i \in \xi} m_i \v_i = m_J \v'_J + \sum_{i \in \xi} m_i \v'_i ,
\end{equation}
for the momentum, where $\xi$ denotes the cell in which the reacting particles
are found and the prime again denotes post-reaction quantities. Summing over all
particles in the cell $\xi$ allows one to utilize the velocities of solvent particles,
other than the reacting ones, to satisfy conservation rules. Energy conservation
reads
\begin{equation}
  \label{eq:en}
  \demi m_J \v_J^2 + \demi \sum_{i \in \xi} m_i \v_i^2 + \sum_{i\in\xi} u_i = \demi m_J \v_J'^2 + \demi \sum_{i \in \xi} m_i \v_i'^2 + \sum_{i\in\xi} u'_i .
\end{equation}
As described below, in order to model endothermic or exothermic reactions we may
associate an internal degree of freedom with the chemical species and in this
equation $u_\alpha$ is the internal energy of species $\alpha$. For
thermoneutral reactions this internal state label is not needed.

Below we shall make use of the following notation: $m_\alpha$ is the mass of
species $\alpha$; $M_{\alpha\beta \ldots \zeta}$ is the sum of the masses of the
different species, for instance, $M_{\alpha\beta} = m_\alpha + m_\beta$ and
$M_{\alpha\alpha}=2m_\alpha$; $\vr_\alpha$ is the position of the particle
$\alpha$; $\vR_{\alpha\beta}$ is the center of mass position of particles
$\alpha$ and $\beta$; $\vrh_{\alpha\beta} =
(\vr_\alpha-\vr_\beta)/|\vr_\alpha-\vr_\beta|$ is the unit vector along the
$\alpha$--$\beta$ direction; $\v_\alpha$ is the velocity of the particle
$\alpha$; $\V_{\alpha\beta}=(m_\alpha\v_\alpha +
m_\beta\v_\beta)/M_{\alpha\beta}$ is the center of mass velocity of $\alpha$ and
$\beta$; $\w_{\alpha\beta} = \v_\alpha - \v_\beta$ is the relative velocity of
$\alpha$ and $\beta$; $\mu_{\alpha\beta} = m_\alpha m_\beta/M_{\alpha\beta}$ is
a reduced mass; and $u_\alpha$ is the internal energy associated to a particle
$\alpha$. For some of the quantities, for instance the sum of the masses or the
center of mass velocity, arbitrary arrangements of indices can be made. For
quantities where the order matters, such as $\w_{\alpha\beta}$ or
$\mu_{\alpha\beta}$, grouping can be made with a parentheses; e.g., $
\mu_{\alpha(\beta\gamma)} = m_\alpha m_{\beta\gamma}/M_{\alpha\beta\gamma}$.

The precise manner in which reaction rules are constructed depends on the
chemical mechanism one wishes to simulate. We illustrate the procedure by
considering two different reaction schemes: an exothermic unimolecular reaction
$A +J \to B+J +\Delta u$, where $\Delta u$ is the energy release on reaction,
and a dissociation reaction $A +J \to J+ 2B$, where $J$ represents the Janus
particle.

\subsection{Exothermic unimolecular reaction}
\label{sec:exo}

The exothermic unimolecular reaction involves the reactant and product species,
$A$ and $B$, respectively, interacting with the Janus particle $J$. The reactive
$A$ and $B$ species, while treated as structureless objects, nevertheless carry
internal energy labels $u_A$ and $u_B$, respectively. For example, this is a
simplified description of a catalyzed isomerization reaction where reactants and
products differ by a free energy difference of $\Delta u=u_B-u_A$.  Mass balance
requires that the masses of $A$ and $B$ be equal: $m_A=m_B$. Recall that the
Janus particle is a rigid object so that no Janus internal degrees of freedom
need to be taken into account. (This condition can be relaxed when flexible
self-propelled particles are considered.)

There is considerable freedom in the way one chooses to implement a reactive
scheme that accounts for the conservation conditions in Eqs.~(\ref{eq:mom}) and
(\ref{eq:en}). For example, for $A+C \to B+C +\Delta u$, we may account for the
energy change $\Delta u$ on reaction by changing the post-collision relative
kinetic energy $\w_{BS_1}'$ of the product $B$ molecule and a solvent $S_1$
molecule, and not the Janus particle. In this case the momentum conservation
equations take the form,
\begin{eqnarray} \label{eq:exo-mom}
m_J \v_J &=& m_J \v'_J \nonumber \\
m_A \v_A + m_{S_1} \v_{S_1} &=& m_B \v'_B + m_{S_1} \v'_{S_1}.
\end{eqnarray}
The second of these equations can be satisfied if the post-collision velocities
are given by
\begin{eqnarray} \label{eq:prime}
  \v'_B &=& \v_A + \frac{\mu_{BS_1}}{m_B} \vs \nonumber \\
  \v'_{S_1} &=& \v_{S_1} - \frac{\mu_{BS_1}}{m_{S_1}} \vs .
\end{eqnarray}
where $\vs$ remains to be determined by energy conservation.

Supposing $u_A > u_B$ for now, we find that the norm of the relative
post-collision velocity $\w_{BJ}'$ depends on the pre-collision relative
velocity $\w_{AJ}$ and also on the difference of internal energies. As $\v_J$
remains unmodified by the reaction, the deposition of the excess energy $\Delta
u$ is made in the relative velocity $\w_{BS_1}'$. Equation~(\ref{eq:en})
becomes, in the center-of-mass velocity frame of $B$-$S_1$,
\begin{equation}
  \label{eq:exo-en1}
  \demi M_{BS_1} \V_{BS_1}^2 + \demi \mu_{BS_1} \w_{BS_1}'^2 = \demi M_{BS_1} \V_{BS_1}^2 + \demi \mu_{BS_1} \w_{BS_1}^2 + u_A - u_B .
\end{equation}
Since the center-of-mass velocity $\V_{BS_1}$ is not altered during the event,
Eq.~(\ref{eq:exo-en1}) becomes after some algebra
\begin{equation}
  \label{eq:exo-en2}
  \w_{BS_1}'^2 = \w_{AS_1}^2 + 2\frac{u_A - u_B}{\mu_{AS_1}}.
\end{equation}
From Eq.~\ref{eq:prime} we have $\w_{BS_1}'$: $ \w_{BS_1}' = \w_{AS_1} + \vs$.  Using
$s=|\vs|$ and $\vsh = \frac{\vs}{s}$ we find
\begin{equation}
  \label{eq:random-extra}
  s = - |\w_{AS_1}| \cos\theta_r \pm \sqrt{\left( |\w_{AS_1}| \cos\theta_r\right)^2 +  2\frac{u_A - u_B}{\mu_{AS_1}}}
\end{equation}
where $\theta_r$ is the angle between $\vs$ and $\w_{BS_1}$. We decide to always pick the $+$ solution. The angle of $\vs$ may be chosen at random.

As a consequence of the local increase in kinetic energy an inhomogeneous
temperature gradient will be generated in the vicinity of the Janus particle. If
the interaction energies of the $A$ and $B$ species with the Janus particle are
same, this model exhibits the interesting feature that propulsion will be driven
by thermophoresis alone.

\subsection{Dissociation reaction} \label{sec:disso}

Here we suppose that a dissociation reaction, $A \to B_1 +B_2$, is catalyzed by
interaction with the rigid Janus particle. The symbols $B_1$ and $B_2$ label the two
$B$ species molecules that result from the dissociation of $A$. Mass
conservation requires that the mass of a $B$ molecules is half that of $A$: $m_B
= m_A/2$. Momentum conservation reads,
\begin{equation}
  \label{eq:disso-mom}
  m_J \v_J + m_A \v_A = m_J \v_J' + m_B ( \v_{B_1}' + \v_{B_2}' ).
\end{equation}
The energy conservation equation, Eq.~(\ref{eq:en}), can be written in the form,
\begin{equation}
  \demi \mu_{AJ} \w_{AJ}^2 =
  \demi \mu_{(B_1B_2)J} \w_{(B_1B_2)J}'^2 + \demi \mu_{B_1B_2} \w_{B_1B_2}'^2.
\end{equation}
We may choose to set $\w_{(B_1B_2)J}'=0$; i.e., we may use all of the relative
kinetic energy of the system to change the relative energy of $B_1B_2$. The
direction of $\w_{B_1B_2}'$ may be chosen at random. The norm of $\w_{B_1B_2}'$
is then given explicitly by
\begin{equation}
   |\w_{B_1B_2}'| = \sqrt{ \frac{\mu_{AJ}}{\mu_{B_1B_2}} \w_{AJ}^2 } ~.
\end{equation}
Other variants of these collision rules may be constructed.

\section{Simulation of Janus particle motion}\label{sec:sim}

In this section we present the results of simulations of the self-propelled
motion of the composite Janus particle when it catalyzes the exothermic $A \to
B$ unimolecular reaction and $A \to 2B$ dissociation reaction. The simulations
were carried out in a cubic box with periodic boundary conditions. This
simulation volume contained $N=N_A+N_B$ molecules. The $A$ and $B$ particles
interact with all beads in the Janus particle through repulsive Lennard-Jones
potentials,
\begin{equation}
  \label{eq:LJ}
  V(r) = 4 \epsilon \left( (\frac{\sigma}{r})^{12} -
    (\frac{\sigma}{r})^{6} + \frac{1}{4} \right) \Theta(r^C - r),
\end{equation}
where $r_c=2^{1/6}\sigma$ and $\Theta$ is the Heaviside function. Note that we
choose the interactions of both species with both catalytic and noncatalytic
beads to be identical; consequently the particles are identical apart from species labels and
no propulsion of the Janus particle can arise from the simple $A \to B$
thermoneutral reaction considered in previous studies of sphere dimer motors. As
described earlier, interactions between the $A$ and $B$ particles in the
environment are accounted for by multiparticle collisions instead of direct
intermolecular forces. In the MD portions of the evolution, Newton's equations
of motion are integrated using the velocity Verlet algorithm with a time step
$\Delta t$. The Shake and Rattle algorithms are used to maintain the rigid bond
constraints in the composite Janus particle. In the MPC portions of the
evolution, grid shifting is applied at each MPCD step to insure Galilean invariance~\cite{ihle_kroll_galilean_invariant_pre_2001}.

All quantities reported below are expressed in simulation units: mass $m_A$,
length $a$, energy $\epsilon$ and time $\sqrt{\frac{m_A a^2}{\epsilon}} \equiv
\tsim$. In these units the system parameters were: simulation box size,
$48\times 48 \times 48$; MPC time $\tau = 0.5$; average number of particles per
MPC cell, 10; reduced temperature, $k_BT=1/3$; and $\Delta t = 0.01$. The choice
of the time step depends on the criterion $\Delta t \ll \tLJ$ where $\tLJ$ is
the natural time for the Lennard-Jones parameters, $\tLJ = \sqrt{\frac{m_A
    \sigma^2}{\epsilon}}$. The Janus particle is made from 36 individual
spheres, each with Lennard-Jones radius $\sigma=3$ and mass $50$, so that $m_J = 1800$. All realizations of the
dynamics started with $100$ MPCD steps with an Andersen
thermostat~\cite{noguchi_et_al_epl_2007} at chosen temperature
$k_BT=1/3$ in systems with all $A$ particles. We note that for these parameters
the thermal velocity for the Janus particle is $ v_{th} = \sqrt\frac{k_BT}{m_J}
\approx 0.014$.

In order to sustain self-propelled motion in the steady state reactants must be
supplied to the system and products removed. In the past this has been done by
introducing fluxes of these species at the boundaries or far from the motor, or
by bulk far-from-equilibrium reactions which are themselves assumed to be sustained
by fluxes. (In simple descriptions these species feeds can be accounted for in
pool chemical species that may be incorporated into effective rate constants.)
In this study we maintain a supply of fresh reagents by taking the reverse of
the reaction on the motor to occur in the bulk of the fluid, but with different
reaction rates. The reverse reaction for the exothermic reaction is $B \to A -
\Delta u$ while for the dissociation reaction it is $2B \to A$. The reaction
rates of both of these reactions were taken to be $k_r=0.001$. These reactions are
implemented using the RMPCD algorithm~\cite{rohlf_et_al_rmpcd_2008}. Momentum
and energy are obeyed at the cell level as in Eqs.~(\ref{eq:mom})
and~(\ref{eq:en}) but without the Janus particle. Energy addition in a cell is
obtained by a velocity scaling in the center-of-mass velocity frame.

\subsection{Exothermic $A \to B$ reaction}

\begin{figure}[htbp]
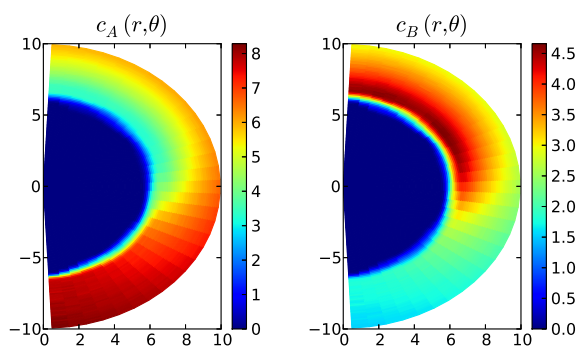

  \centering
  \includegraphics[width=\linewidth]{{{JEX_K3_U0.5_1_C_pf}}}
  \caption{The concentrations field for $A$ and $B$ molecules, $c_A$ and $c_B$ for $\Delta u=0.5$.}
  \label{fig:exo-pf}
\end{figure}

\begin{figure}[htbp]
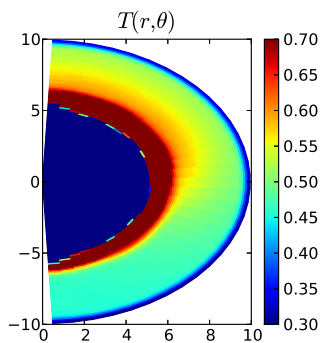

  \centering
  \includegraphics[width=\myw\linewidth]{{{JEX_K3_U0.5_1_C_temp}}}
  \caption{The temperature field around the Janus particle for $\Delta u=0.5$. The black region is related to the very low solvent
    concentrations $c_\alpha(r,\theta)$ close to the Janus particle that leads to very poor statistics.}
  \label{fig:exo-temp}
\end{figure}

As noted earlier, since the $A$ and $B$ particles are identical apart from species labels,
the only possible source of directed motion (assuming instabilities do not
occur) is the temperature gradient across the Janus particle generated by the
exothermic reaction. In simulations starting from all $A$ molecules in the
simulation box, after a transient of $3000$ time units, the combined effect of
reactions at the Janus particle and in the bulk leads to a steady average value
of $N_A$ and $N_B$. For the system in this steady state regime, the number
concentrations $c_A$ and $c_B$ are displayed in Fig.~\ref{fig:exo-pf} and show a
strong excess of $B$ at the top (catalytic side) of the particle. The polar axis of the Janus particle is used to define a coordinate system that
moves along the particle. The definitions of the radius $r$ and angle $\theta$
are shown in Fig.~\ref{fig:coord} (right). The
temperature field $T(r,\theta)$ around the compound is shown in
Fig.~\ref{fig:exo-temp}. The temperature is computed cell-wise at regular
intervals during the simulations and binned at the center of mass of the
corresponding cell. One clearly observes a difference between the $C$ and $N$
sides of the Janus particle, a sign that a thermal gradient is present in the
stationary regime of the system.

The velocity of the self-propelled Janus particle can be determined from the
time and ensemble average of $V_z(t)$, the instantaneous velocity of the
center-of-mass velocity projected on the instantaneous unit vector from the
center of mass $N$ hemisphere to the center of mass of the $C$ hemisphere,
\begin{equation}
  \label{eq:vzt}
  \langle V_z\rangle(t)  = \Big\langle \frac{1}{t} \int_0^{t} dt'~ \vr_{CN}(t') \cdot \v_J(t')\Big\rangle,
\end{equation}
where the angular brackets denote an average over realizations of the
evolution. When the time argument is omitted, $\langle V_z\rangle=\langle V_z\rangle(t_{max})$ where $t_{max}$
is the duration of the simulation.

\begin{figure}[htbp]
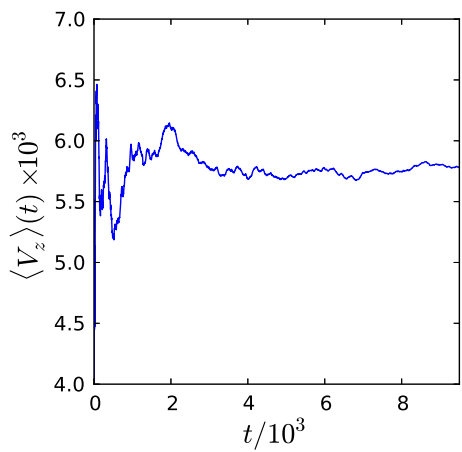

  \centering
  \includegraphics[width=0.8\columnwidth]{{{JEX_K3_U0.5_l_vzt}}}
  \caption{The running average of the directed velocity for $\Delta u = 0.5$. The result is obtained from an average of 16 independent realizations of the dynamics.}
  \label{fig:JEX}
\end{figure}

Figure~\ref{fig:JEX} displays $\langle V_z \rangle (t)$. After a short
transient, the running average stabilizes around an approximately constant nonzero value,
indicating the existence of directed motion. The value of the velocity is
positive, consistent with the results
obtained by Yang and Ripoll~\cite{yang_ripoll_thermophoretic_pre_2011} where the
thermal gradient is generated externally on the fluid around one sphere of a
sphere dimer. The directed velocity $V_z$ is not negligible compared to the
thermal velocity $v_{th}$ and the directed component of the motion plays an
important role. Thus, we have demonstrated that the chemically induced
temperature gradient by the chemical reaction leads to self-thermophoretic
propulsion of the Janus motor.

The behavior for different values of $\Delta u$ is summarized in
Table~\ref{tab:exo-vz} where $\langle V_z \rangle$ and its standard deviation, which
characterizes the coherence of $V_z$ between different runs of the same
physical conditions, are given. As expected, increasing the energy $\Delta u$ associated to
the reaction increases the propulsion velocity.
\begin{table}[htbp]
  \centering
  \begin{tabular}{l | l l l l}
    & $\Delta u = 0.1$ & $\Delta u = 0.2$ & $\Delta u = 0.3$ & $\Delta u = 0.5$ \\
    \hline
    \hline
    $\langle V_z \rangle \times 10^{3}$ & $1.5$ & $2.8$ & $3.8$ & $5.7$ \\
    $\sigma_{\langle V_z \rangle} \times 10^{3}$ & $0.8$ & $0.9$ & $1.2$ & $1.4$
  \end{tabular}
  \caption{The average directed velocity and the associated standard deviation for the exothermic reaction for several values of the exothermicity. Results are averages over $16$ independent realizations.}
  \label{tab:exo-vz}
\end{table}

\subsection{Dissociation $A \to 2B$ reaction}

\begin{figure}[htbp]
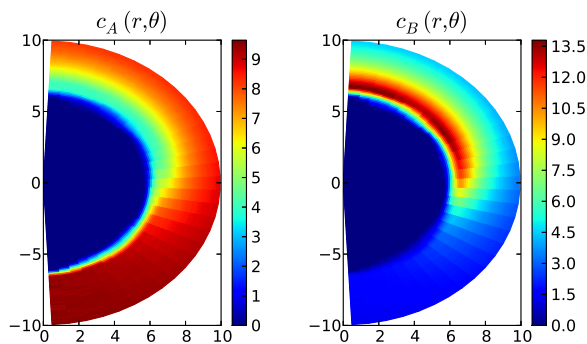

  \centering
  \includegraphics[width=\linewidth]{{{ABC_17_K3_1_F_pf}}}
  \caption{The concentration fields, $c_\alpha(r,\theta)$, for $A$ and $B$ molecules for the dissociation reaction..}
  \label{fig:disso-pf}
\end{figure}

A similar set of simulations was carried out for the dissociation
reaction. Figure~\ref{fig:disso-pf} displays the concentration fields
$c_\alpha(r,\theta)$ for the $A$ and $B$ species, showing the production of $B$
on the top catalytic side of the Janus particle. As each $A$ produces $2B$, the
maximum value of $c_B$ is much higher than the average density of $A$ particles
in the system, about $13.5$ particles per MPC cell. The $B$ particles diffuse
from the $C$ side of the Janus particle to the bulk of the fluid where the
reverse reaction takes place.

\begin{figure}[htbp]
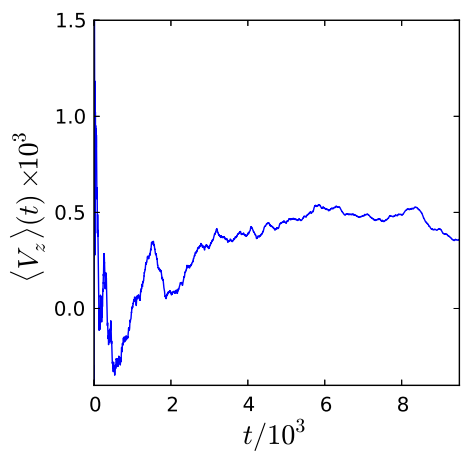

  \centering
  \includegraphics[width=0.8\columnwidth]{{{ABC_17_K3_l_vzt}}}
  \caption{The running average of the directed velocity for the dissociation reaction.}
  \label{fig:ABC}
\end{figure}

Figure~\ref{fig:ABC} shows $\langle V_z \rangle(t)$ [from
Eq.~(\ref{eq:vzt})]. The gradient of $c_A$ and $c_B$ results in directed motion with the $C$ hemisphere of the Janus particle in front.
The value of $V_z$, however, is much lower than the thermal
velocity $v_{th}$ of the Janus particles and the propulsive character can only
be observed through averaging.
Thus, even though the $A$ and $B$ particles interact with beads of the Janus particle with the same intermolecular potentials, the
excess of $B$ particles produced in the reaction provides a mechanism for self-diffusiophoresis and propulsion is observed.

\section{Conclusions}\label{sec:conc}

The directed motions of small nano- and micron-scale motors that operate by
chemically-fueled phoretic mechanisms depend on the nature of the catalytic
chemical reactions that occur on the motor. These reactions generate the
gradients of fields that are responsible for the propulsion of the motor. The
work described in this investigation demonstrated how a variety of chemical
mechanisms can be incorporated into a mesoscopic description of the dynamics of
the system. One of the important aspects of propulsion by phoretic mechanisms is
the correct description of the fluid flow fields that are induced by the motor
motion. In order to properly describe the hydrodynamics of the motor
environment, the mesoscopic dynamics must preserve the basic conservation laws
of mass, momentum and energy conservation. The hybrid MD-MPCD dynamics employed
in this study satisfies these criteria. We have shown how different catalytic
reactions at the motor surface can be constructed that maintain these
conservation laws.

The two examples we have chosen to illustrate the method present
interesting features. For both the exothermic unimolecular and
dissociation reactions we have chosen to consider the situation where
the $A$ and $B$ reactive species interact with the same intermolecular
forces with the Janus particle. This is an special but interesting
case that highlights the role that the reaction mechanism plays in the
motor propulsion. In the case of the exothermic reaction, the net
effect of the reaction is to generate an inhomogeneous temperature
gradient in the vicinity of the Janus particle, giving rise to
propulsion by a thermophoretic mechanism. This scheme is easily
generalized to allow for different interaction potentials for the
chemical species on the different hemispheres of the Janus
particle. In such a situation both thermophoretic and diffusiophoretic
mechanisms will operate, and they can act in either the same or opposite directions. For the dissociation reaction, it is the
increased number of product molecules produced in reaction, even when
the potential energies of interaction of reactant and products with
the Janus particle are the same, that leads to propulsion by
self-diffusiophoresis. Again if these potential energies were
different the propagation velocity would change.

The work presented here provides a method that allows one to explore
self propulsion for a variety of motor geometries and catalytic
reaction mechanisms at a mesoscopic level. Through such studies one
can investigate motor dynamics on scales where the validity continuum
theories should be tested, and study situations where analytical methods
are difficult to carry out due to the complex nature of the reaction dynamics.

{\it Acknowledgments}: Research supported in part by a grant from the
Natural Sciences and Engineering Research Council of
Canada. Computations were performed on the GPC supercomputer at the
SciNet HPC Consortium. SciNet is funded by: the Canada Foundation for
Innovation under the auspices of Compute Canada; the Government of
Ontario; Ontario Research Fund - Research Excellence; and the
University of Toronto.

\footnotesize{
\bibliography{janus-chem-v2} 
\bibliographystyle{rsc} 
}

\end{document}